\documentclass[12pt,preprint]{aastex}
 \begin{document}

\title{Fundamental vibration mode in a highly inhomogeneous star}

\author{
S.I. Bastrukov\altaffilmark{1,3}, H.-K. Chang\altaffilmark{1,2}, E.-H. Wu\altaffilmark{2}, I.V. Molodtsova\altaffilmark{3}}

\altaffiltext{1}{Institute of Astronomy,\\
  National Tsing Hua University, Hsinchu, 30013, Taiwan}

\altaffiltext{2}{Department of Physics,\\
  National Tsing Hua University, Hsinchu, 30013, Taiwan}

\altaffiltext{3}{Laboratory of Informational Technologies,\\
Joint Institute for Nuclear Research, 141980 Dubna, Russia}

\begin{abstract}
   The eigenfrequency problem of fundamental vibration mode in a highly inhomogeneous star modeled by self-gravitating mass of viscous liquid with singular density at the center is considered
   in juxtaposition with that for Kelvin fundamental mode in the liquid star model with
   uniform density. Particular attention is given to the difference between spectral equations for the
   frequency and lifetime of $f$-mode in the singular and homogeneous star models.
   The newly obtained results are discussed in the context of theoretical asteroseismology of white dwarfs.
\end{abstract}

\section{Introduction}
  The term fundamental mode has been introduced in the theory of stellar pulsations by Cowling (1941)
  from hydrodynamical model of a heavy mass of inviscid incompressible liquid of uniform density $\rho$ undergoing free oscillations with nodeless irrotational velocity of fluctuating flow.
  The extended discussion of the eigenfrequency problem of this model, whose solution is due to Kelvin (1863), can be found elsewhere (e.g. Lamb 1945, Chandrasekhar 1961, Bastrukov 1996). An outstanding importance of this fiducial model for theoretical asteroseismology is that it
  sets the standard for analytic study of non-radial pulsations of the main-sequence stars (Cox 1980, Unno et al 1989) and serves as an example, according to Chandrasekhar (1961), of "at least one problem for which analytic solution can be found" and which illustrates "the type of difficulties one must confront in the other problem", because "in most instances, the problems become of such complexity and involve so many parameters that elementary methods of solution seem impracticable".

  The purpose of this work is to explore some peculiarities of the fundamental vibration mode in the inhomogeneous stars. In approaching the topic like this it seems better to start with a brief outline of
  assumptions lying at the base of the $f$-mode frequency computation in the homogeneous liquid star model in which the uniform equilibrium density is not altered, $\delta \rho=0$. The obvious consequence of this assumption steaming from the continuity equation
  \begin{eqnarray}
 \label{e1.1}
 \delta{\dot \rho}=-\nabla \,(\rho\,\delta {\bf v})=-(\delta {\bf v}\cdot\nabla)\,\rho-\rho\,
 \nabla\cdot \delta {\bf v}
 \end{eqnarray}
 is that the velocity of flow oscillating in a spherical mass of incompressible liquid of uniform density is described by the potential vector field of the form
  \begin{eqnarray}
  \label{e1.2}
  && \nabla\cdot  \delta {\bf v}({\bf r},t)=0,\quad
  \delta {\bf v}({\bf r},t)=\nabla \chi({\bf r},t),\\
  \label{e1.3}
  && \nabla^2 \chi({\bf r},t)=0,\quad
  \chi({\bf r},t)=A_\ell\,r^\ell\,P_\ell(\cos\theta){\dot \alpha}(t).
  \end{eqnarray}
 The last equation exhibits nodeless character of the velocity flow as a function of distance from center to the surface of the star. It is this feature of oscillating flow is regarded as the major kinematic
 signature of fundamental vibration mode.

 In this paper we relax the basic assumption of the homogeneous liquid star model about uniform equilibrium density, preserving the above major kinematic signature of $f$-mode.
 Specifically, we consider admittedly idealized a highly inhomogeneous liquid star model
 in which the non-uniform density profile has singularity at the star center of the form
\begin{eqnarray}
 \label{e1.4}
 \rho(r)=\frac{5}{6}\rho_s \left({\frac{R}{r}}\right)^{1/2}.
 \end{eqnarray}
 One of the most conspicuous features of this density profile
 is that the total mass of such a star
 \begin{eqnarray}
 \label{e1.5}
 M=\int \rho(r)\,d{\cal V}= \frac{4\pi}{3}\rho_s R^3
 \end{eqnarray}
 is finite and identical to that for the total mass of the canonical homogenous liquid star model.
 In somewhat different context the model of inhomogeneous liquid star with similar density profile has been briefly discussed by Clayton (1986).
 This curious feature of the model, which from now on is referred to as the singular star model,
 is interesting in its own right because it permits too analytically tractable solution of the eigenfrequency problem for $f$-mode. Perhaps the most striking feature distinguishing
 vibrational behavior of inhomogeneous from homogeneous one is that the nodeless non-rotational vibrations of a spherical liquid mass of non-uniform density are of substantially compressional character, as it follows from the continuity equation. Therefore, one of the prime purposes of our study
 is to find out how this striking distinction between the density profiles of homogeneous and inhomogeneous models is reflected in the frequency and lifetime spectra of fundamental vibration mode.

 The paper is organized as follows. In section 2, a general mathematical treatment of the
 inhomogeneous star vibrations in the fundamental mode on the basis of the Rayleigh's energy
 variational principle is outlined.
 In section 3, the detailed analytic computation of the spectral equations for the frequency and lifetime of $f$-mode in the singular liquid star model is presented followed by their comparison with those for the Kelvin fundamental mode in homogeneous star model which is the subject of Section 4.
 The newly obtained results are highlighted in section 5 and
 briefly discussed in the context of current development of theoretical  asteroseismology of white dwarfs.

\section{General equations of nodeless stellar vibrations in fundamental mode}

 The state of motion of flowing stellar matter under the action of forces of
 buoyancy, gravity and viscous stress is uniquely described in terms of five dynamical variables, to wit, the density $\rho({\bf r},t)$, three components
 of the flow velocity ${\bf v}({\bf r},t)$ and the pressure $p({\bf r},t)$ obeying the coupled equations of fluid-mechanics and Newtonian universal gravity
 \begin{eqnarray}
 \label{e2.1}
 &&\frac{d\rho}{dt}=-\rho\,
 \frac{\partial v_k}{\partial x_k},\quad\quad \frac{d}{dt}=
 \frac{\partial }{\partial t}+v_k\frac{\partial }{\partial x_k},
 \\[0.2cm]
 \label{e2.2}
 && \rho\frac{dv_i}{dt}=-
 \frac{\partial p}{\partial x_i}+\rho\frac{\partial U}{\partial x_i}+
 \frac{\partial \pi_{ik}}{\partial x_k},
 \\[0.2cm]
 \label{e2.3}
 && \frac{dp}{dt}=\frac{\Gamma_1 p}{\rho}\frac{d\rho}{dt}=-\Gamma_1\,p\frac{\partial v_k}{\partial x_k}, \quad \quad \Gamma_1=\frac{\partial \ln p}{\partial \ln \rho},
  \\[0.2cm]
 \label{e2.4}
 &&\nabla^2 U=-4\pi G\rho.
 \end{eqnarray}
 The equation for density $\rho$ is the condition of continuity of flowing stellar matter. The Navier-Stokes equation for the velocity flow expresses the second law of  Newtonian dynamics
 for viscous liquid and equation for pressure is the condition of adiabatic behavior of stellar matter; $\Gamma_1$ stands for the adiabatic coefficient of gaseous pressure in the star. This latter
 condition means that in the process of motion, like vibrations, the time scale of energy exchange between infinitesimally close elementary volume of liquid is much longer than characteristic period of oscillations (e.g. Huang 1987, Kawaler \& Hansen 1994).
 The tensor of Newtonian viscous stresses is given by
 \begin{eqnarray}
  \label{e2.5}
  \pi_{ik}=2\eta\,v_{ik}+\left(\zeta-\frac{2}{3}\,\eta\right)v_{jj}\,\delta_{ik},\quad\quad
  v_{ik}=\frac{1}{2}\left(\nabla_k v_i+\nabla_i v_k\right).
  \end{eqnarray}
  In computing frequency of fundamental vibration mode, the equilibrium density profile
  $\rho$, the pressure $P_c$ at the star center as well as the transport coefficients of stellar
  matter, the shear $\eta$ and the bulk $\zeta$ viscosities, are regarded
  as input, in advance given, parameters.

  The potential of self-gravity $U(r)$ and pressure $p(r)$ in motionless, ${\bf v}=0$,
  state of hydrostatic equilibrium, are the solutions of coupled equations
  \begin{eqnarray}
 \label{e2.6}
 \nabla_i\,p(r)=\rho(r)\nabla_i\, U(r),\quad\quad \nabla^2 U(r)=-4\pi G\rho(r).
 \end{eqnarray}
 The gravity potential $U$ is determined, in effect, by Poisson equation for internal
 potential $U_i$ and Laplace equation for external one $U_e$
 supplemented by the standard boundary conditions of the continuity of these potentials and their normal
 derivatives on the star surface
 \begin{eqnarray}
 \label{e2.7}
 && \nabla^2 U_i(r)= - 4\pi G \rho(r),\quad r\leq R,\quad
 \quad\nabla^2 U_e(r)=0,\quad r>R,\\[0.2cm]
\label{e2.8}
 && U_i(r)=U_e(r)\Big\vert_{r=R},\quad\quad
 \frac{dU_i(r)}{dr}=\frac{dU_e(r)}{dr}\Big\vert_{r=R}.
 \end{eqnarray}
 The general solution of equation for pressure is specified by standard boundary condition of stress-free surface $p(r=R)=0$.

 The equations of linear oscillations of stellar matter about
 stationary state of hydrostatic equilibrium of a star with non-uniform equilibrium density, $\rho=\rho(r)$, are obtained by applying to (\ref{e2.1})-(\ref{e2.4}) the standard procedure of linearization
 \begin{eqnarray}
 \label{e2.9}
 && \rho\to\rho(r)+\delta \rho({\bf r},t)\,\quad  p\to p(r)+\delta p({\bf r},t)\\
 \label{e2.10}
 && v_i\to v_i+\delta v_i({\bf r},t),\quad [v_i=0],\quad U\to U ({\bf r})+\delta U({\bf r},t).
 \end{eqnarray}
 As was stated, in this work we focus on the regime of irrotational vibrations in which the velocity field
 of fluctuating flow subjects to
 \begin{eqnarray}
 \label{e2.11}
 \nabla\cdot \delta {\bf v}=0,\quad  \nabla\times \delta {\bf v}=0.
 \end{eqnarray}
 Given this, the continuity equation takes form
 \begin{eqnarray}
 \label{e2.12}
 \delta{\dot \rho}=-(\delta v_k\nabla_k)\rho.
 \end{eqnarray}
 The liner fluctuations of the flow velocity are governed by linearized Navier-Stokes equation
 \begin{eqnarray}
 \label{e2.13}
 &&\rho\delta {\dot v}_i=-
 \nabla_i\delta p+\rho\nabla_i\delta U+\delta\rho\nabla_i U
 +\nabla_k \delta\pi_{ik},
 \\[0.2cm]
 \label{e2.14}
 &&\delta \pi_{ik}=2\eta \delta v_{ik},\quad\quad
 \delta v_{ik}=\frac{1}{2}(\nabla_i\delta v_k+\nabla_k\delta v_i).
 \end{eqnarray}
 From (\ref{e2.3}) it follows that rate of change in the pressure is controlled by equation
 \begin{eqnarray}
 \label{e2.15}
 \delta {\dot p}=-(v_k\nabla_k)\,p.
 \end{eqnarray}
 Fluctuations in potential of self-gravity inside the star $\delta U=\delta U_i$, caused by fluctuations in density $\delta \rho$, subject to the Poisson equation
  \begin{eqnarray}
 \label{e2.16}
 \nabla^2 \delta U=-4\pi G\,\delta \rho.
 \end{eqnarray}
 The energy balance in the process of oscillations is controlled by equation
 \begin{eqnarray}
 \label{e2.17}
 &&\frac{\partial }{\partial t}\int \frac{\rho\delta v^2}{2}\,d{\cal V}=
 -\int [(\delta v_k\nabla_k)\delta p - \rho (\delta v_k\nabla_k) \delta U
 - \delta \rho (\delta v_k\nabla_k) U-\delta\pi_{ik}\delta v_{ik}]\,
  d{\cal V}
 \end{eqnarray}
 which is obtained after scalar multiplication of (\ref{e2.13}) by $\delta v_i$
 and integration over the star volume.
 To compute the eigenfrequency of vibrations we take advantage of the Rayleigh's energy variational principle at the base of which lies the following separable representation of fluctuating
 variables
 \begin{eqnarray}
 \label{e2.18}
 && \delta v_i({\bf r},t)=a_i({\bf r}){\dot \alpha}(t),\quad \delta U({\bf r},t)=\phi({\bf r}){\alpha}(t),\\
 \label{e2.19}
 && \delta \rho({\bf r},t)=\tilde \rho({\bf r})\,{\alpha}(t),\quad \tilde \rho({\bf r})=-(a_k({\bf r})\nabla_k)\,\rho(r),\\
 \label{e2.20}
 &&\delta p({\bf r},t)=\tilde p({\bf r})\,{\alpha}(t),\quad \tilde p({\bf r})=-(a_k({\bf r})\nabla_k)\,p(r),\\
 \label{e2.21}
 && \delta \pi_{ik}({\bf r},t)=2\eta\,\delta v_{ik}=2\eta\,a_{ik}({\bf r}){\dot \alpha}(t),\\
 \label{e2.22}
 && \delta v_{ik}({\bf r},t)=a_{ik}({\bf r}){\dot \alpha}(t),
 \quad\quad a_{ik}({\bf r})=\frac{1}{2}(\nabla_ia_k({\bf r}),
 +\nabla_k a_i({\bf r})).
 \end{eqnarray}
 Hereafter $a_i({\bf r})$ stands for the time-independent field of instantaneous material
 displacements and $\alpha(t)$ for the temporal amplitude of oscillations.
 The key idea of such representation is that it transforms equation of energy
 balance (\ref{e2.17}) into equation for $\alpha$  having well-familiar form
 of equation of damped oscillations
  \begin{eqnarray}
 \label{e2.23}
 && \frac{d{\cal E}}{dt}=-2{\cal F},\quad\quad
 {\cal E}=\frac{{\cal M}{\dot\alpha}^2}{2}+\frac{{\cal K}\alpha^2}{2},\quad
 {\cal F}=\frac{{\cal D}{\dot \alpha^2}}{2},
 \\[0.2cm]
 \label{e2.24}
 && {\cal M}{\ddot \alpha}+{\cal D}{\dot \alpha}+{\cal K}\alpha=0,
 \\[0.2cm]
 \label{e2.25}
 && {\cal M}=\int \rho(r) a_k \,a_k\,d{\cal V},\quad {\cal D}=2\int \eta(r) a_{ik}\,a_{ik}\,d{\cal V}\\
 && {\cal K}=\int [(a_k\,\nabla_k\,\rho(r))\, (a_k\,\nabla_k\,U({\bf r}))-\rho(r) (a_k\,\nabla_k)\phi({\bf r})-(a_k\nabla_k)\,(a_k\nabla_k) p(r)]\,d{\cal V}.
 \end{eqnarray}
 Here ${\cal E}$ is the energy of free, non-dissipative, oscillations and ${\cal F}$ is the dissipative
 function of Rayleigh describing their damping by shear viscosity of stellar matter.
 The solution of (\ref{e2.24}) is given by
 \begin{eqnarray}
 \label{e2.26}
 && \alpha(t)=\alpha_0\,\exp(-t/\tau)\,\cos \omega(\tau) t,
 \\[0.2cm]
 \label{e2.27}
 && \omega^2(\tau)=\omega^2[1-(\omega\tau)^{-2}],
 \quad\omega^2=\frac{{\cal K}}{{\cal M}},\quad
  \tau=\frac{2{\cal M}}{{\cal D}}
 \end{eqnarray}
  where $\omega(\tau)$ is the frequency of dissipative oscillations damped by viscosity, $\omega$ is the
 frequency of free oscillations and $\tau$ is their lifetime. Thus, to compute
 the frequency and lifetime one need to specify all variables entering integral parameters of inertia
 ${\cal M}$, stiffness ${\cal K}$ and viscous friction ${\cal D}$.

 We start
 with the field of instantaneous displacements $a_k({\bf r})$ which is
 the key kinematic characteristics of $f$-mode.
 In the star undergoing irrotational oscillations in the $f$-mode, the shape of an arbitrary spherical surface takes the form of harmonic spheroids which are described by
 \begin{eqnarray}
 \label{e2.28}
 r(t) = r [ 1+ \alpha_\ell (t) P_\ell (\zeta)],\quad \zeta=\cos\theta
\end{eqnarray}
 where $P_\ell (\zeta)$ is the Legendre polynomial of the multipole order $\ell$ specifying
 the overtone number in fundamental vibration mode.
 In the system with fixed polar axis the potential field of velocity is found from Laplace equation
 supplemented by boundary condition that radial component of velocity on the star surface equals
 to the rate of the surface distortions taking the shape of harmonic spheroids
 \begin{eqnarray}
 \label{e2.29}
 && \nabla_k\,\delta v_k=0,\quad \delta v_k=\nabla_k \chi,\quad \nabla^2 \chi=0\quad\to\quad \chi=A_\ell\, r^\ell\, P_\ell(\zeta)\,{\dot\alpha}(t),\\
 \label{e2.30}
 && \delta v_r\big\vert_{r=R}={\dot R}(t),\quad  R(t) = R [ 1+ \alpha_\ell(t) P_\ell (\zeta)].
\end{eqnarray}
Taking into account that $\delta {\bf v}({\bf r},t)={\bf a}({\bf r}){\dot\alpha}(t)$ one has
 \begin{eqnarray}
 \label{e2.31}
 {\bf a}({\bf r})=A_\ell\nabla r^\ell\,P_\ell(\zeta),\quad\quad
 A_\ell=\frac{1}{\ell R^{\ell-2}}.
\end{eqnarray}
In terms of this field, the linearized equations for the density $\delta \rho$ and for the
pressure $\delta p$ reads
\begin{eqnarray}
 \label{e2.32}
 && \delta \rho({\bf r},t)=\tilde \rho({\bf r})\,{\alpha}(t),\quad  \tilde \rho({\bf r})=-({\bf a}
 \cdot \nabla)\,\rho(r)=-A_\ell P_\ell(\zeta) \ell\,r^{\ell-1}\frac{\partial \rho(r)}{\partial r},\\
 \label{e2.33}
 && \delta p({\bf r},t)=\tilde p({\bf r})\,{\alpha}(t),\quad  \tilde p({\bf r})
 =-({\bf a}\cdot \nabla)\,p(r)=-A_\ell P_\ell(\zeta) \ell\,r^{\ell-1}\frac{\partial p(r)}{\partial r}
 \end{eqnarray}
where $\rho(r)$ and $p(r)$ are the density and pressure of gravitationally equilibrium, hydrostatic, configuration.

 To compute variations in the potential of gravity one must consider
 two equations for internal $\delta U_i$ and external $\delta U_i$ potentials
 \begin{eqnarray}
 \label{e2.34}
 && \nabla^2 \delta U_i({\bf r},t)=-4\pi\,G\delta \rho({\bf r},t),\quad  \nabla^2 \phi_i({\bf r})=-4\pi\,G \tilde \rho({\bf r})\\
  \label{e2.35}
 && \nabla^2 \delta U_e({\bf r},t)=0,\quad  \nabla^2 \phi_e({\bf r})=0.
 \end{eqnarray}
The outlined energy variational method provides a general framework for computing
the frequency of $f$-mode in a inhomogeneous Newtonian liquid star with arbitrary form of
non-uniform density profile.

\section{The singular star model}

 As was stated, we use the term singular star for a self-gravitating mass of viscous
 liquid with the non-uniform and singular in the star center density profile given
 by equation (\ref{e1.4}) whose total mass is identical to that for homogenous star model.
 For such a singular star, the equilibrium potentials and fields of universal gravity are given by
 \begin{eqnarray}
  \label{e3.1}
 && \nabla^2 U_i(r)=-4\pi G \rho(r),\quad \rho(r)=\frac{5}{6}\rho_s \left({\frac{R}{r}}\right)^{1/2},
 \quad M=\int \rho(r)\,d{\cal V}= \frac{4\pi}{3}\rho_s R^3,\\
 \label{e3.2}
 && U_i(r<R)=\frac{20\pi}{9}G\rho_s R^2\left[1-\frac{2}{5}
 \left(\frac{r}{R}\right)^{3/2}\right]=\frac{5}{3}\frac{G M}{R}\left[1-\frac{2}{5}
 \left(\frac{r}{R}\right)^{3/2}\right],\\
 \nonumber
 &&{\rm \bf g}_i(r<R)=-\nabla U_i=
 \frac{G M}{R^3}\left(\frac{r}{R}\right)^{1/2}{\bf r},\\[0.2cm]
 \label{e3.3}
 && U_e(r>R)=\frac{4\pi}{3}G\rho_s \frac{R^3}{r}=\frac{G M}{r},\quad {\rm \bf g}_e(r>R)=-\nabla U_e=\frac{G M}{r^3} {\bf r}.
 \end{eqnarray}
 The hydrostatic pressure obeying the boundary condition of free surface $p(r)\vert_{r=R}=0$ reads
 \begin{eqnarray}
 \label{e3.4}
 p(r)=\frac{10\pi}{9}G\rho_s^2R\left(R-r\right)=P_c\left(1-\frac{r}{R}\right),\quad P_c=\frac{10\pi}{9}G\rho_s^2R^2=\frac{5}{8\pi}\frac{G M^2}{R^4}.
 \end{eqnarray}
 It is remarkable that the pressure at the star center, where the density has singularity,
 is finite and its radial profile is linear function of distance from the center to the star surface. The pressure in the center is defined by equation of state of the stellar matter.\footnote{In white dwarfs the central pressure is identified with the pressure of degenerate Fermi-gas of ultra relativistic electrons: $P_c=P_F=K\,n_e^{4/3}$. The remarkable feature of the emergence of
 these end products of stellar evolution (e.g. Hansen \& Kawaler 1994) is that the density at the center of a white dwarf tends to infinity as the mass of collapsing star approaches the Chandrasekhar limiting mass (e.g. Phillips 1994).}. With in advance given equation of state for $P_c$ the rightmost identity
 in (\ref{e3.4}) is considered as definition of the star radius. The internal gravitational energy is
 \begin{eqnarray}
 \label{e3.5}
 W_s=\frac{1}{2}\int \rho_s\, U_i d{\cal V}=\frac{11}{18}\frac{GM^2}{R}\simeq 1.02 W,\quad W=\frac{3}{5}\frac{GM^2}{R}
  \end{eqnarray}
where $W$ is the total gravitational energy of homogeneous star of equivalent mass $M$.
To take into account the compression effect of self-gravity on mechanical property of star matter we adopt the suggestion of work (Bastrukov et al 2007a) that radial profile of viscosity coefficient is identical to that for the equilibrium pressure profile, that is, of the form
\begin{eqnarray}
 \label{e3.6}
 \nu(r)=\nu_c\left(1-\frac{r}{R}\right)
 \end{eqnarray}
where $\nu_c$ is the shear viscosity in the star center which along with $\rho_s$ and
$P_c$ are regarded as input parameters of the model.

\subsection{Exact solution of Poisson equation for variations of self-gravity potential in singular star}

 Having defined the equilibrium profiles of density $\rho(r)$, the pressure $p(r)$, the shear viscosity profile $\nu(r)$ and knowing the field of instantaneous displacements
 ${\bf a}({\bf r})$ we are able to compute the inertia ${\cal M}$ and viscous friction ${\cal D}$.
 However, in order to compute the stiffness ${\cal K}$ we must calculate fluctuations in the potential
 of self-gravity $\delta U({\bf r},t)=\phi({\bf r})\alpha(t)$, that is, to solve Poisson equation
 for $\phi({\bf r})$ with a fairly non-trivial right part
  \begin{eqnarray}
 \label{e3.7}
 &&\nabla^2 \phi_i(r,\zeta)=-\frac{5\pi}{3} A_{\ell}\, G \rho_s\, R^{1/2}\,\ell\, r^{\ell-5/2} P_{\ell}(\zeta).
 \end{eqnarray}
 In the spherical polar coordinates we have
 \begin{eqnarray}
 \label{e3.8}
 && \frac{1}{r^2}\frac{\partial }{\partial r}r^2\frac{\partial \phi_i(r,\theta)}{\partial r}+\frac{1}{r^2\sin\theta}\frac{\partial }{\partial \theta}\sin\theta\frac{\partial \phi_i(r,\theta)}{\partial \theta}=-\frac{5\pi}{3}\,A_{\ell}\, G \rho_s\, R^{1/2}\,\ell\, r^{\ell-5/2} P_{\ell}(\cos\theta).
 \end{eqnarray}
 Assuming a solution of the form
 \begin{eqnarray}
 \label{e3.9}
 && \phi_i(r,\theta)=\frac{u(r)}{r}\,P_\ell(\cos\theta)
 \end{eqnarray}
 and taking into account that $P_\ell(\cos\theta)$ is the solution of equation
 \begin{eqnarray}
 \label{e3.10}
 && \frac{1}{\sin\theta}\frac{\partial }{\partial \theta}\sin\theta\frac{\partial P_\ell(\cos\theta)}{\partial \theta}=-\ell(\ell+1)\,P_\ell(\cos\theta)
 \end{eqnarray}
 we obtain
 \begin{eqnarray}
 \label{e3.11}
 && r^2\frac{\partial^2 u(r)}{\partial r^2}-\ell(\ell+1)u(r)=-\frac{5\pi}{3}\,A_\ell\, G \rho_s\, R^{1/2}\,\ell\, r^{\ell+1/2}.
 \end{eqnarray}
 The general solution of equation for $\phi_i$, which is finite at the origin, is given by
\begin{eqnarray}
\label{e3.12}
&& \phi_i(r,\theta)=\left\{C\,r^\ell+\frac{20\pi}{3}\frac{\ell}
{4\ell+1}\,\,A_\ell\,\rho_s\, G\,R^{1/2}r^{\ell-1/2}\right\}P_\ell(\cos\theta)
 \end{eqnarray}
 Outside the star we have
  \begin{eqnarray}
  \label{e3.13}
&&\nabla^2 \delta U_e=0\quad\to\quad \nabla^2 \phi_e({\bf r})=0,\\
\label{e3.14}
&& \phi_e=D\, r^{-\ell -1}P_\ell(\cos\theta).
 \end{eqnarray}
 The arbitrary constants $C$ and $D$ are eliminated from boundary conditions
 \begin{eqnarray}
 \label{e3.15}
\phi_i=\phi_e\Big\vert_{r=R},\quad\quad \frac{\partial \phi_i}{\partial r}=\frac{\partial \phi_e}{\partial r}\Big\vert_{r=R}.
\end{eqnarray}
which yield
 \begin{eqnarray}
 \label{e3.16}
 C=-\frac{10\pi}{3}\,A_\ell\,\rho_s\,G\,\frac{\ell}{2\ell+1},\quad
 D=\frac{10\pi}{3}\,A_\ell\, \rho_s\, G\, \frac{\ell}{(4\ell+1)(2\ell+1)}R^{2\ell+1}.
 \end{eqnarray}
Finally, we obtain
\begin{eqnarray}
\label{e3.17}
&& \phi_i(r,\theta)=-\frac{10\pi}{3}\frac{\ell}{2\ell+1}A_\ell\, \rho_s\, G\,\left[r^\ell
-\frac{2(2\ell+1)}{4\ell+1}R^{1/2}
\,r^{\ell-1/2}\right]P_\ell(\cos\theta).
 \end{eqnarray}
It is worth emphasizing that following this line of argument one can get the solutions
of Poisson equation for a more wide class of inhomogeneous star models undergoing nodeless spheroidal vibrations with non-rotational field of velocity. Also, the method and obtained solution can be useful
in the study of electrodynamic problems of astrophysical interest.

\subsection{Spectral equations for frequency and lifetime of $f$-mode}

The mass parameter ${\cal M}$ is given by
\begin{eqnarray}
\nonumber
{\cal M}&=&\int_V \rho(r)\,[a^2_r+a^2_\theta]\,d{\cal V}
 = \frac{5\pi}{3}\rho_s A_\ell^2 R^{1/2}\int\limits_{0}^{R}\,r^{2\ell-1/2}\,dr\,\int\limits_{-1}^{+1}
\left[\ell^2 P^2_\ell(\zeta)+\left(1-\zeta^2\right)\left(\frac{dP_{\ell}(\zeta)}{d\zeta}\right)^2\right] d\zeta\\
\label{e3.18}
&=& \frac{20\pi}{3}\,A_\ell^2\,\rho_s\,R^{2\ell +1}\,\frac{\ell}{4\ell +1}.
\end{eqnarray}
Computation of the viscous friction parameter, with
the non-uniform profile of shear viscosity (\ref{e3.6}), yields
\begin{eqnarray}
\nonumber
 {\cal D}&=& 2 \int \eta(r) a_{ij}a_{ij}\, d{\cal V}
 = 2 \int\eta(r)\left( a_{rr}^2+a_{\theta\theta}^2+a_{\phi\phi}^2+2 a_{r\theta}^2\right) d{\cal V}
 \\ [0.2cm]\nonumber
 &=&  8\pi A_\ell^2\int\limits_{0}^{R}  \eta(r) r^{2\ell-2} dr \int\limits_{-1}^{1}
  \left[
 \ell^2 (\ell^2 -\ell+1) P_\ell(\zeta)^2
 -\ell (\ell+1) \zeta P_\ell(\zeta)\frac{dP_\ell(\zeta)}{d\zeta}+
 \zeta^2\left(\frac{dP_\ell(\zeta)}{d\zeta}\right)^{2}\right. \\ \nonumber
 &+&  \left.
 (\ell-1)^2(1-\zeta^2)\left(\frac{d P_\ell(\zeta)}{d\zeta}\right)^{2}\right]
 d\zeta =8\pi A_\ell^2\,\ell(\ell-1)(2\ell-1)\,\int\limits_{0}^{R}
 \eta(r)\, r^{2\ell-2} dr\\
 \label{e3.19}
 &=& 4\pi
 A_\ell^2\,\eta_c\, R^{2\ell-1}(\ell-1).
\end{eqnarray}
The lifetime, $\tau=2{\cal M}/{\cal D}$, of $\ell$-pole overtone of $f$-mode is given by
\begin{eqnarray}
 \label{e3.20}
 && \tau_f(\ell)=\frac{10}{3}\tau_\nu \frac{\ell}{(4\ell+1)(\ell-1)},\quad \tau_\nu=\frac{R^2}{\nu},\quad\nu=\frac{\eta_c}{\rho_s}.
 \end{eqnarray}
The integral parameter of stiffness
 \begin{eqnarray}
 \label{e3.21}
 && {\cal K}=\int [(a_k\,\nabla_k\,\rho(r))\, (a_k\,\nabla_k\,U({\bf r}))-(a_k\nabla_k)\,(a_k\nabla_k) p(r)-\rho(r) (a_k\,\nabla_k)\phi({\bf r})]\,d{\cal V}
 \end{eqnarray}
can be conveniently represented in the form ${\cal K}=K_1+K_2+K_3$, where
\begin{eqnarray}
 \nonumber
  K_1&=&\int (a_k\,\nabla_k\,\rho(r))\, (a_k\,\nabla_k\,U({r}))\,d{\cal V}=
 \int \left(a_r(r,\theta)\,\frac{\partial \rho(r)}{\partial r}\right)\,
 \left(a_r(r,\theta)\,\frac{\partial U(r)}{\partial r}\right)\,d{\cal V}\\
 \label{e3.22}
 &=&\frac{10\pi^2}{9}G\rho_s^2A_\ell^2\,R\,\ell^2
 \int\limits_{0}^{R} r^{2\ell-1}dr\,\int\limits_{-1}^{+1} P_\ell^2(\zeta)d\zeta=
 \frac{10\pi^2}{9}G\rho_s^2\,A_\ell^2\,R^{2\ell+1}\,\frac{\ell}{2\ell+1}.
 \end{eqnarray}
The integral for $K_2$ is given by
\begin{eqnarray}
\nonumber
 K_2&=&-\int(a_k\nabla_k)\,(a_k\nabla_k) p(r)\,d{\cal V}=-\int
 \left[a_r(r,\theta)\,\frac{\partial }{\partial r}+\frac{a_\theta(r,\theta)}{r}\frac{\partial }{\partial \theta} \right]\left(a_r(r,\theta)\,\frac{\partial p(r)}{\partial r}\right)\,d{\cal V} \\
 \nonumber
 &=&\frac{10\pi^2}{9}G\rho_s^2R\int\limits_{0}^{R}r^{2\ell-1}dr\,
  \left[\ell^2(\ell-1)\int\limits_{-1}^{+1}P_\ell^2(\zeta)d\zeta+
  \ell\int\limits_{-1}^{+1}(1-\zeta^2)^{1/2}\left(\frac{dP_\ell(\zeta)}{d\zeta}\right)^2d\zeta\right]\\
  \label{e3.23}
  &=&\frac{40\pi^2}{9}A_\ell^2G\rho_s^2R^{2\ell+1}\frac{\ell^2}{2\ell+1}.
 \end{eqnarray}
In similar fashion, for $K_3$ we obtain
\begin{eqnarray}
  \nonumber
 K_3&=&-\int \rho(r) (a_k\,\nabla_k)\phi({\bf r})\,d{\cal V}=-\int \rho(r)
 \left[a_r(r,\theta)\,\frac{\partial \phi(r,\theta)}{\partial r}+\frac{a_\theta(r,\theta)}{r}\frac{\partial \phi(r,\theta)}{\partial \theta}\right]\,d{\cal V}\\
 \nonumber
 &=&\frac{-50\pi^2}{9} G \rho_0^2 A_{\ell}^2 R
\left[\int\limits_{0}^{R}\left(\frac{\ell ^3 r^{2\ell-5/2}}{\left(2\ell+1\right)R^{1/2}}-\ell ^2\frac{2\ell-1}{4\ell+1}r^{2\ell-3}\right)r^2 dr\int\limits_{-1}^1P_{\ell}^2(\zeta)\left(\zeta\right)d\zeta
\right. \\ \nonumber
 & & \qquad\qquad\qquad +  \left.
\int\limits_{0}^{R} \left(\frac{\ell r^{2\ell -5/2}}{\left(2\ell +1\right)R^{1/2}}-\frac{2\ell r^{2\ell-3}}{4\ell+1}\right)r^2 dr\int\limits_{-1}^1\left(1-\zeta^2\right)\left(\frac{dP_{\ell}(\zeta)}{d\zeta}\right)^2 d\zeta\right] \\
\label{e3.24}
 &=&-\frac{50\pi^2}{9}A_\ell^2G\rho_s^2R^{2\ell+1}\frac{\ell}{(2\ell+1)(4\ell+1)}.
 \end{eqnarray}
The resultant expression for stiffness ${\cal K}$ reads
\begin{eqnarray}
\label{e3.25}
{\cal K}=\frac{40\pi^2}{9}A_\ell^2G\rho_s^2R^{2\ell+1}\frac{\ell(4\ell^2+2\ell-1)}{(2\ell+1)(4\ell+1)}.
 \end{eqnarray}
 From analytic form of ${\cal M}$ and ${\cal K}$ it follows that in the singular star model under consideration the lowest overtone of $f$-mode is of dipole degree, $\ell=1$.

 \begin{figure}[h]
 \centering{\includegraphics[width=11cm]{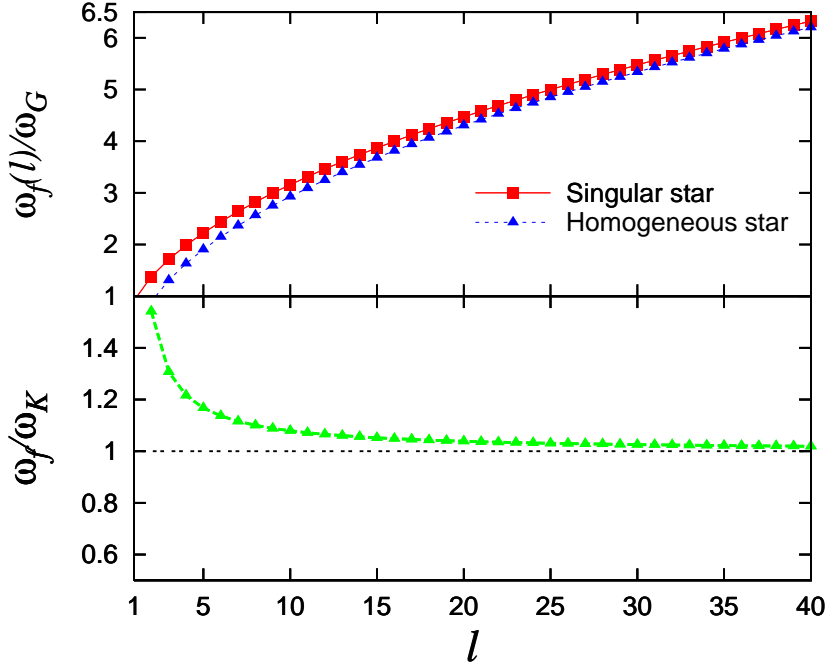}}
 \caption{Frequency of fundamental modes $\omega_f$ as a function of multipole degree $l$ of global nodeless irrotational vibrations of singular and homogeneous star normalized to
 the natural unit of frequency $\omega_G$ (upper panel), and the ratio of spectral equations for the frequency of fundamental mode in singular model $\omega_f$ and in the
 Kelvin homogeneous model $\omega_K$.}
 \end{figure}

 The frequency spectrum of fundamental vibration mode in the singular star model
 reads
\begin{eqnarray}
\label{e3.26}
\omega_f^2( _0g_\ell)=\omega^2_G\frac{2\ell(2\ell+1)-1}{2(2\ell+1)},\quad\quad \omega^2_G=\frac{4\pi}{3}\,G\,\rho_s=\frac{GM}{R^3},\quad\quad \ell\geq 1.
 \end{eqnarray}
This last equation can be recast in the following equivalent form
\begin{eqnarray}
\label{e3.27}
&&\omega_f( _0g_\ell)=\omega_G\left[\ell-\frac{1}{2(2\ell+1)}\right]^{1/2},\\
&&\ell >> 1 \quad\quad \omega_f(_0g_\ell)\simeq\omega_G\sqrt{\ell}.
 \end{eqnarray}
 showing asymptotic behavior of the frequency at very high overtones.

 The obtained frequency spectrum of $f$-mode in singular star model has many features in common with
 the well-known Kelvin spectral formula for kinematically identical fundamental vibration mode in homogeneous star model
 \begin{eqnarray}
 \label{e3.28}
\omega_K=\omega_f( _0g_\ell)=\omega_G\left[\frac{2\ell(\ell-1)}{(2\ell+1)}\right]^{1/2},\quad\quad\omega_G=\sqrt{\frac{GM}{R^3}},\quad \quad\ell\geq 2
 \end{eqnarray}
 but the lowest overtone of Kelvin $f$-mode is of quadrupole degree, $\ell=2$.
 As is demonstrated in Fig.1, the most essential differences between the above spectral equations
 are manifested in low-overtone domain (upper panel) and that at large $\ell$ the frequency spectra of $f$-mode in singular and homogeneous star models shear identical asymptotic behavior (lower panel).
 In the next section this last spectral formula is briefly recovered by the above expounded
 method with allow for the effect of viscous damping of $f$-mode whose lifetime is computed with non-uniform profile of shear viscosity
 identical in appearance to that for the profile of hydrostatic pressure in this model.

\section{Kelvin $f$-mode in the homogeneous liquid star}

 In the canonical homogeneous star model of uniform density, $\rho={\rm constant}$
 the total mass has one and the same form as in above singular model, $M=(4\pi/3)\rho R^3$.
 The internal and external
 potentials of Newtonian gravitational field are the solutions of Poisson equation inside and Laplace equation outside the star
 \begin{eqnarray}
 \label{e4.1}
 && \nabla^2 U_i=-4\pi\,G\rho:\,\, U_i(r<R) = 2\pi G\rho R^2\left[1-\frac{1}{3}\left(\frac{r}{R}\right)^2\right]=
 \frac{3}{2}\frac{G M}{R}\left[1-\frac{1}{3}\left(\frac{r}{R}\right)^2\right], \\
 \label{e4.2}
 && W=\frac{1}{2}\int \rho\, U_i\, d{\cal V}=\frac{3}{5}\frac{GM^2}{R},\quad \nabla^2 U_e=0:\,\,  U_e(r>R)=\frac{4\pi}{3}G\rho \frac{R^3}{r}=\frac{G M}{r},\\
   \label{e4.3}
 &&{\rm \bf g}=-\nabla U:\quad {\rm \bf g}_i(r<R)=\frac{G M}{R^3}\,{\bf r},\quad\quad
 {\rm \bf g}_e(r>R)=\frac{G\,M}{r^3}\,{\bf r}.
  \end{eqnarray}
 In (\ref{e4.2}), $W$ is the internal gravitational energy $W$ of homogeneous star.
 The solution of equation of hydrostatic for pressure $p(r)$ is given by
 \begin{eqnarray}
 \label{e4.4}
 && \nabla p(r)=\rho\nabla U_i(r)\quad\to\quad
 p(r)=\frac{2\pi}{3}G\rho^2(R^2-r^2)=P_c\left[1-\left(\frac{r}{R}\right)^2\right],\\
 \label{e4.5}
 && P_c=\frac{2\pi}{3}G\rho^2R^2=\frac{3}{8\pi}\frac{G M^2}{R^4},\quad R=\sqrt{\frac{3P_c}{{2\pi}{G\rho^2}}}.
 \end{eqnarray}
 The last expression shows again that the star radius is determined by the central pressure $P_c$ related to the density $\rho$ by equation of state.

 \begin{figure}[h]
 \centering{ \includegraphics[width=6.5cm]{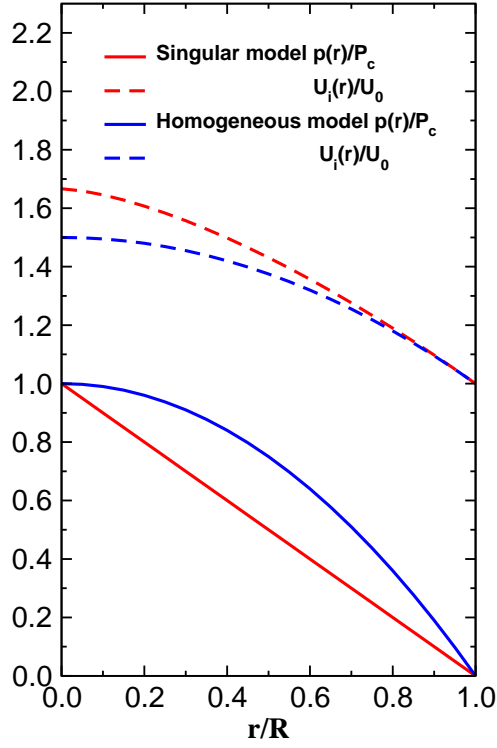}}
 \caption{
 The fractional pressure and gravity potential profiles in singular inhomogeneous star model
 and in the canonical homogenous star model.}
\end{figure}

 For comparison, in Fig.2 we plot the fractional pressure and gravity potential profiles
 computed in homogeneous model
\begin{eqnarray}
\label{e4.6}
 \frac{p(r)}{P_c}=\left[1-\left(\frac{r}{R}\right)^2\right],\quad  P_c=\frac{3}{8\pi}\frac{G M^2}{R^4},\quad  \frac{U_i(r)}{U_0}=\frac{3}{2}\left[1-\frac{1}{3}
 \left(\frac{r}{R}\right)^{2}\right],\quad U_0=\frac{G M}{R}
 \end{eqnarray}
  and in singular inhomogeneous model
 \begin{eqnarray}
 \label{e4.7}
 \frac{p(r)}{P_c}=\left(1-\frac{r}{R}\right),\quad P_c=\frac{5}{8\pi}\frac{G M^2}{R^4},\quad
 \frac{U_i(r)}{U_0}=\frac{5}{3}\left[1-\frac{2}{5}
 \left(\frac{r}{R}\right)^{3/2}\right],\quad U_0=\frac{G M}{R}.
 \end{eqnarray}
 The governing equations for non-compressional irrotational vibrations of the homogeneous liquid star are
 \begin{eqnarray}
 \label{e4.8}
 &&\rho\delta {\dot v}_i=-
 \nabla_i\delta p+\rho\nabla_i\delta U+\nabla_k \delta\pi_{ik},\\
 \label{e4.9}
 &&\delta {\dot p}=-(v_k\nabla_k)\,p,\quad \nabla^2 \delta U=0,\\
 \label{e4.10}
 && \delta \pi_{ik}=2\eta \delta v_{ik},\quad
 \delta v_{ik}=\frac{1}{2}(\nabla_i\delta v_k+\nabla_k\delta v_i),\quad \eta=\eta_c\left[1-\left(\frac{r}{R}\right)^2\right].
 \end{eqnarray}
 The equation of energy conservation is
 \begin{eqnarray}
 \label{e4.11}
 &&\frac{\partial }{\partial t}\int \frac{\rho\delta v^2}{2}\,d{\cal V}=
 -\int [(\delta v_k\nabla_k)\delta p - \rho (\delta v_k\nabla_k) \delta U
 -\delta\pi_{ik}\delta v_{ik}]\,
  d{\cal V}.
 \end{eqnarray}
 On substituting here
 $$\delta v_i({\bf r},t)=a_i({\bf r}){\dot \alpha} (t),\quad
 \delta U({\bf r},t)=\phi_i({\bf r}){\alpha}(t),\quad \delta p({\bf r},t)=-(a_k\nabla_k)\,p(r){\alpha}(t) $$
 we arrive at equation of damped oscillator, ${\cal M}{\ddot \alpha}+{\cal D}{\dot \alpha}+{\cal K}\alpha=0$, with integral parameters defining the frequency and lifetime
 of the form
  \begin{eqnarray}
 \label{e4.12}
 {\cal M}&=&\int_{\cal V}\rho\, a_k \,a_k\,d{\cal V},\quad
 {\cal D}=2\int \eta(r) a_{ik}\,a_{ik}\,d{\cal V},\\
 \label{e4.13}
 {\cal K}&=&-\int_{\cal V} [(a_i\nabla_i)\,(a_k\nabla_k) p(r)+\rho\,(a_k\,\nabla_k)\phi({\bf r})]\,d{\cal V},\\
 && \label{e4.14}
 \omega^2=\frac{{\cal K}}{{\cal M}},\quad
  \tau=\frac{2{\cal M}}{{\cal D}}.
 \end{eqnarray}
 The only unknown quantity is the variations of the gravity potentials obeying the
 Laplace equations
 \begin{eqnarray}
 \label{e4.15}
 \nabla^2 \delta U_i=0,\quad\quad \nabla^2 \delta U_e=0
 \end{eqnarray}
 having the general solutions of the form
 \begin{eqnarray}
 \label{e4.16}
 \delta U_i(r<R)=C_\ell\,r^\ell P_\ell(\zeta)\alpha(t), \quad\quad
 \delta U_e(r<R)=D_\ell\,r^{-(\ell+1)}P_\ell(\zeta)\alpha(t).
 \end{eqnarray}
 The arbitrary constants $C_\ell$ and $D_\ell$ are eliminated from
 the standard boundary conditions
 \begin{eqnarray}
 \label{e4.17}
 && U_i(r(t))+\delta U_i(r(t))=U_e(r(t))+\delta U_e(r(t))\Big\vert_{R=R},
 \\[0.2cm]
 \label{e4.18}
 && \frac{d}{dr}
 \left[U_i(r(t))+\delta U_i(r(t))\right]=
 \frac{d}{dr}\left[U_e(r(t))+\delta U_e(r(t))\right]\Big\vert_{r=R}
 \end{eqnarray}
 where $r(t)=r[1+ \alpha(t) P_\ell (\cos\theta)]$. Retaining in these equations terms of first order in $\alpha(t)$ and putting $r=R$ we arrive at coupled algebraic equations for $C_\ell$ and $D_\ell$ whose solution leads to the following final expressions for the time-independent part of gravity potential (Bastrukov 1996a)
 \begin{eqnarray}
 \label{e4.19}
&&  \phi_i=\frac{4\pi}{2l+1}\frac{G\rho}{R^{\ell-2}}
 \,r^\ell\,P_\ell(\zeta),\quad
  \phi_e=\frac{4\pi}{2\ell+1}G\rho\,R^{\ell+3}
 \,r^{-(\ell+1)}P_\ell(\zeta).
 \end{eqnarray}
Computation of integral parameters of the inertia ${\cal M}$ viscous friction ${\cal D}$ yields
 \begin{eqnarray}
 \label{e4.20}
 && {\cal M}=4\pi\,A^2_\ell\,\rho R^{2\ell+1}\frac{\ell}{2\ell+1}=\frac{4\pi\rho R^5}{\ell\left(2\ell +1\right)},
 \\[0.2cm]
 \label{e4.21}
 && {\cal D}=16\pi\,A_\ell^2\eta_c R^{2\ell-1}\frac{\ell(\ell-1)}{2\ell+1}=16\pi \eta_c R^3\frac{(\ell-1)}{2\ell+1}.
 \end{eqnarray}
 Following the suggestion of Cowling (1941), one can consider separately oscillations restored by force associated
 with gradient in fluctuations of pressure and force owing its origin to fluctuations in the potential of
 gravity. In accord with this, the integral parameter of stiffness is written  represents as a sum
\begin{eqnarray}
\label{e4.22}
&& {\cal K}={\cal K}_p+{\cal K}_g,\\
\label{e4.23}
&& {\cal K}_p=-\int_{\cal V}(a_k\nabla _k)\,(a_i\nabla _i)p(r)\,d{\cal V},\quad {\cal K}_g=-\int_{\cal V}
 \rho(a_i\nabla _i)\phi({\bf r})\,d{\cal V}
\end{eqnarray}
and, consequently, the frequency frequency of vibrations is also represented as a sum
\begin{eqnarray}
\label{e4.24}
&& \omega^2=\omega_p^2+\omega_g^2,\quad\omega_p^2=\frac{{\cal K}_p}{\cal M},\quad \omega_g^2=\frac{{\cal K}_g}{\cal M}
\end{eqnarray}
where $\omega_p$ designate the frequency of $p$-mode and $\omega_g$ is the frequency of $g$-mode.
For ${\cal K}_p$ and ${\cal K}_g$ we get
\begin{eqnarray}
\label{e4.25}
{\cal K}_p=\frac{16\pi^2}{3}\frac{G\rho^2R^5}{2\ell+1},\quad {\cal K}_g=-\frac{16\pi^2 G\rho^2 R^5}{(2\ell+1)^2}.
\end{eqnarray}
For the frequencies of $p$-mode and $g$ mode we obtain
\begin{eqnarray}
\label{e4.26}
&& \omega_p^2=\frac{{\cal K}_p}{\cal M}=\omega_G^2\,\ell,\quad\quad
\omega_g^2=\frac{{\cal K}_g}{\cal M}=-\omega_G^2\frac{3\ell}{2\ell+1}.
\end{eqnarray}
One sees that $\omega_p^2$ is positive whereas $\omega_g^2$ is negative\footnote{These sings for $p$-mode and $g$-mode are one and the same as in well-known dispersion relation of Jeans $\omega^2=c_s^2k^2-4\pi G\rho$ which characterizes
propagation of sound, with velocity $c_s$, in a self-gravitating fluid of constant density $\rho$ (e.g. Chandrasekhar 1961, Sect. 119)}.
 It follows that frequency of fundamental mode $\omega_f(_0g_\ell)=[\omega_p^2+\omega_g^2]^{1/2}$ is given by Kelvin spectral formula (\ref{e3.28}) and for the lifetime of $\ell$-pole overtone, $\tau_f(\ell)$, computed with the
 non-uniform radial profile of shear viscosity $\eta=\eta(r)$ given by equation (\ref{e4.10}), we obtain
\begin{eqnarray}
 \label{e4.27}
 && \tau_f(\ell)=\frac{\tau_\nu}{2(l-1)},\quad\quad
 \tau_\nu=\frac{R^2}{\nu},\quad \nu=\frac{\eta_c}{\rho}.
 \end{eqnarray}

 \begin{figure}[h]
 \centering{ \includegraphics[width=11cm]{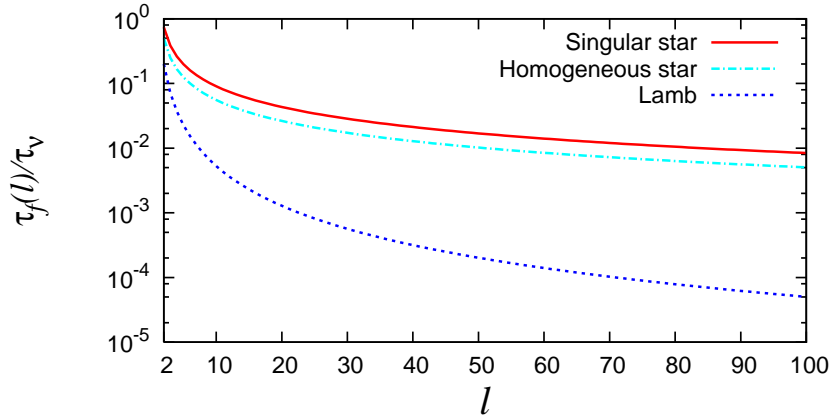}}
 \caption{
 The fractional lifetime of $\ell$-pole overtones of fundamental vibrational mode in homogeneous and inhomogeneous liquid star models with non-uniform profiles of shear viscosity in juxtaposition
 with Lamb spectral equation for damping time of nodeless irrotational oscillations of a spherical mass
 of viscous liquid of uniform density and constant coefficient of shear viscosity.}
\end{figure}

It is worthwhile to compare the computed lifetime spectra with the Lamb (1945) spectral equation
\begin{eqnarray}
 \label{e4.28}
 && \tau_{\rm Lamb}(\ell)=\frac{\tau_\nu}{(2\ell+1)(l-1)}
 \end{eqnarray}
 that has been obtained in a similar fashion but assuming that coefficient of shear viscosity has
 constant value in the entire spherical volume of homogeneous viscous liquid (see also Bastrukov et al 2007).
 It has been pointed by Jeffreys (1976), however, in the context of geoseismology that
 the approximation of uniform viscosity
 does not allows for the effect of self-gravity on mechanical property of matter of astrophysical objects.
 With this in mind, we have supposed that it would be not inconsistent to take radial
 profile of shear viscosity
 similar to that for hydrostatic pressure in the star. Also noteworthy is that Newtonian law of shear viscosity is equally appropriate
 for viscous liquid and viscoelastic solid (e.g. Landau et al 1986). The obtained here spectral
 equations for the time of viscose damping of nodeless spheroidal vibrations may be of some interest,
 therefore, for the general seismology of Earth-like planet (Aki \& Richards 2002).
 In Fig.3, the obtained spectral equations for the lifetime, normalized to $\tau_\nu$, as a function of overtone number $\ell$ are plotted for both singular (\ref{e3.20}) and homogeneous (\ref{e4.27}) models
 in juxtaposition with the Lamb spectral formula (\ref{e4.28}).

\section{Summary}

 The most striking differences between vibrational behavior of homogenous and inhomogeneous liquid star models in the fundamental mode of global nodeless irrotational pulsations
 under the combined action of the buoyancy and gravity forces is that in the inhomogeneous model the oscillations are of substantially compressional character, that is, accompanied by fluctuations in density and the lowest overtone of $f$-mode is of dipole degree, as has first been observed in (Podgainy et al 1996). In the homogeneous star model
 they are characterized as non-compressional and the lowest overtone is of quadrupole degree.

  \begin{figure}[h]
 \centering{\includegraphics[width=8cm]{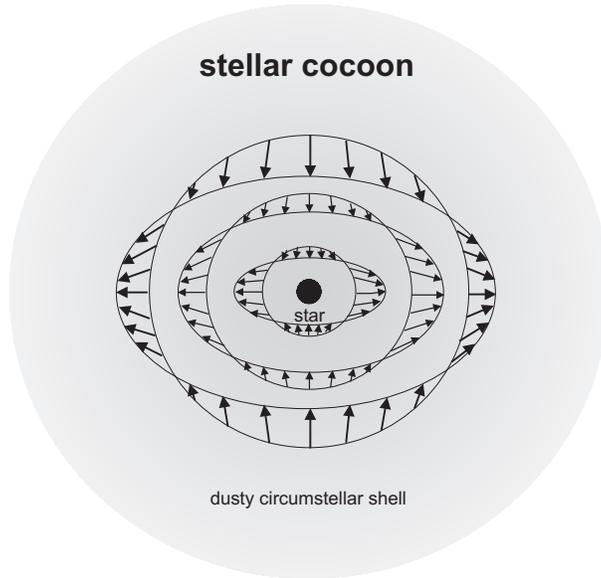}}
 \caption{Artist view of a spherical star-forming object, the stellar cocoon, whose
 largescale vibrations in $f$-mode can be analyzed on the basis considered highly inhomogeneous model.}
 \end{figure}

 The natural unit of frequency of $f$-mode, $\omega_G=\sqrt{GM/R^3}$, has one and the same form as
 that for gravity or $g$-modes. These letter modes are of particular importance for the asteroseismology
 of transitory pre-white dwarfs objects and young white dwarfs (e.g. Hansen \& Kawaler 1994) whose
 variability of electromagnetic emission is attributed to non-radial
 gravity-driven seismic vibrations (Koester \& Chanmugam 1990, Fontaine \& Brassard 2008, Winget  \& Kepler 2008). To this end, it is worth emphasizing, the existing modal classification of the $g$-mode spectra presumes that white-dwarf-forming object oscillates in the standing-wave regime in which the frequency is determined by nodal structure of fluctuating variables (e.g. Unno et al 1989, Winget  \& Kepler 2008). According to standard nomenclature, the $g$-modes of standing-wave regime of non-radial pulsations are specified as $_ng_\ell$, where $n\geq 1$ is the node number and $\ell$ is the multipole degree of overtone. In the meantime, the regime of nodeless oscillations, designated as $_0g_\ell$,
 remains less studied. Also, the considered model can be invoked in assessing  variability of emission from star-forming clouds like stellar cocoon, pictured in Fig.4, as caused by its vibrations in fundamental mode. Motivated by these
 arguments we have investigated an analytic model of a highly inhomogeneous star undergoing
 global nodeless irrotational vibrations. In work (Bastrukov 1996b) these kind of nodeless oscillations have been studied in the model of homogeneous liquid layer. In recent papers (Bastrukov et al 2007b, 2008), the nodeless vibrations of a solid star driven by restoring force of elastic deformations and
 locked in the peripheral finite-depth mantle, has been been studied in the context of
 seismic vibrations of the neutron star crust in the approximation of uniform density of crustal matter. Understandably that in real white dwarfs and neutron stars
 the bulk density is non-homogeneous function of position. However, this realistic case of nodeless irrotational vibrations entrapped in the peripheral finite-depth seismogenic layer with non-uniform density profile demands more elaborate mathematical treatment which will be the subject of our forthcoming paper.

\acknowledgments

 The authors are grateful to Gwan-Ting Chen (NTHU, Hsinchu, Taiwan) for helpful assistance. This work is a part of projects on investigation of variability of high-energy emission from compact sources
 supported by NSC of Taiwan, grant numbers  NSC-96-2628-M-007-012-MY3 and NSC-97-2811-M-007-003.

\end{document}